\documentclass[twocolumn]{aastex62}
\pdfoutput=1 
\usepackage{amsmath,amstext}
\usepackage[T1]{fontenc}
\usepackage[figure,figure*]{hypcap}
\usepackage{rotating}
\usepackage{lineno}
\usepackage{hyperref}
\usepackage[shortlabels]{enumitem}
\usepackage{color}
\usepackage{array}
\usepackage{xspace}
\usepackage{multirow}
\usepackage{acronym}
\graphicspath{{./}{figures/}}

\def\GtrTwoHundredMarginalizedGRBOne{\ensuremath{78\%}\xspace}
\def\GtrTwoHundredMarginalizedGRBTwo{\ensuremath{91\%}\xspace}

\def\DistanceTraveled{\ensuremath{895~\mathrm{kpc}}\xspace}
\def\PostSNSystemic{\ensuremath{135~\mathrm{km\,s}^{-1}}\xspace}

\def\PercentileGRBOne{\ensuremath{\gtrsim\,4\%}\xspace}
\def\PercentileGRBTwo{\ensuremath{98^{\rm th}}\xspace}
\def\PercentileFiftyGRBOne{\ensuremath{\approx\,2.5\%}\xspace}
\def\PercentileFiftyGRBTwo{\ensuremath{\approx\,3.2\%}\xspace}
\def\FiveReffApprox{\ensuremath{\approx\,10\%}\xspace}
\def\TenReffApprox{\ensuremath{\approx\,3\%}\xspace}

\def\NTracersFiveGyr{\ensuremath{<\,4\%}\xspace}
\def\MassAfterQuenching{\ensuremath{\lesssim\,1.4\%}\xspace}
\def\TQuenchGRBOne{\ensuremath{1.6~\mathrm{Gyr}}\xspace}
\def\TQuenchGRBTwo{\ensuremath{3.6~\mathrm{Gyr}}\xspace}

\def\VsysNinteyLowDMGRBOne{\ensuremath{180~\mathrm{km\,s}^{-1}}\xspace}
\def\VsysNinteyLowDMGRBTwo{\ensuremath{230~\mathrm{km\,s}^{-1}}\xspace}
\def\VsysNinteyHighDMGRBOne{\ensuremath{305~\mathrm{km\,s}^{-1}}\xspace}
\def\VsysNinteyHighDMGRBTwo{\ensuremath{445~\mathrm{km\,s}^{-1}}\xspace}

\def\TinspMedGRBOne{\ensuremath{\sim\,2.7~\mathrm{Gyr}}\xspace}
\def\TinspMedGRBTwo{\ensuremath{\sim\,5.0~\mathrm{Gyr}}\xspace}

\def\LowKickPercentageTwoSigamGRBTwo{\ensuremath{0.2\%}\xspace}

\def\VsysNinetyGRBOne{\ensuremath{\gtrsim\,180-305~\mathrm{km\,s}^{-1}}\xspace}
\def\VsysNinetyGRBTwo{\ensuremath{\gtrsim\,230-445~\mathrm{km\,s}^{-1}}\xspace}
\def\LargerOffsetsThanGRBOne{\ensuremath{\simeq\,4\%}\xspace}
\def\LargerOffsetsThanGRBTwo{\ensuremath{\simeq\,2\%}\xspace}

\acrodef{GRB}{gamma-ray burst}
\acrodef{SN}{supernova}
\acrodefplural{SN}[SNe]{supernovae}
\acrodef{DNS}{double neutron star}
\acrodef{SFH}{star formation history}
\acrodef{SFMS}{star-forming main sequence}
\acrodef{SFR}{star formation rate}
\acrodef{SMHM}{stellar mass--halo mass}
\acrodef{DM}{dark matter}
\acrodef{NFW}{Navarro--Frenk--White}

\definecolor{chmagenta}{rgb}{0.54, 0.17, 0.88}
\definecolor{lukegreen}{rgb}{0.0, 0.4, 0.03}

\shorttitle{Progenitors of Highly-Offset Short \acp{GRB}}
\shortauthors{Zevin et al.}

\begin{document}

\title{Forward Modeling of Double Neutron Stars: \\Insights from Highly-Offset Short Gamma-Ray Bursts}

\author[0000-0002-0147-0835]{Michael Zevin}\thanks{NASA Hubble Fellow}\thanks{michaelzevin@uchicago.edu}
\affiliation{Department of Physics and Astronomy, Northwestern University, 2145 Sheridan Road, Evanston, IL 60208, USA}
\affiliation{Center for Interdisciplinary Exploration and Research in Astrophysics (CIERA), 1800 Sherman Avenue, Evanston, IL 60201, USA}
\affiliation{Enrico Fermi Institute and Kavli Institute for Cosmological Physics, The University of Chicago, 5640 South Ellis Avenue, Chicago, Illinois 60637, USA}

\author[0000-0002-6625-6450]{Luke Zoltan Kelley}
\affiliation{Department of Physics and Astronomy, Northwestern University, 2145 Sheridan Road, Evanston, IL 60208, USA}
\affiliation{Center for Interdisciplinary Exploration and Research in Astrophysics (CIERA), 1800 Sherman Avenue, Evanston, IL 60201, USA}

\author[0000-0002-2028-9329]{Anya Nugent}
\affiliation{Department of Physics and Astronomy, Northwestern University, 2145 Sheridan Road, Evanston, IL 60208, USA}
\affiliation{Center for Interdisciplinary Exploration and Research in Astrophysics (CIERA), 1800 Sherman Avenue, Evanston, IL 60201, USA}

\author[0000-0002-7374-935X]{Wen-fai Fong}
\affiliation{Department of Physics and Astronomy, Northwestern University, 2145 Sheridan Road, Evanston, IL 60208, USA}
\affiliation{Center for Interdisciplinary Exploration and Research in Astrophysics (CIERA), 1800 Sherman Avenue, Evanston, IL 60201, USA}

\author[0000-0003-3870-7215]{Christopher P.\ L.\ Berry}
\affiliation{Department of Physics and Astronomy, Northwestern University, 2145 Sheridan Road, Evanston, IL 60208, USA}
\affiliation{Center for Interdisciplinary Exploration and Research in Astrophysics (CIERA), 1800 Sherman Avenue, Evanston, IL 60201, USA}

\author[0000-0001-9236-5469]{Vicky Kalogera}
\affiliation{Department of Physics and Astronomy, Northwestern University, 2145 Sheridan Road, Evanston, IL 60208, USA}
\affiliation{Center for Interdisciplinary Exploration and Research in Astrophysics (CIERA), 1800 Sherman Avenue, Evanston, IL 60201, USA}
\affiliation{CIFAR Fellow}

\begin{abstract}

We present a detailed analysis of two well-localized, highly offset short gamma-ray bursts---GRB~070809 and GRB~090515---investigating the kinematic evolution of their progenitors from compact object formation until merger. 
Calibrating to observations of their most probable host galaxies, we construct semi-analytic galactic models that account for star formation history and galaxy growth over time. 
We pair detailed kinematic evolution with compact binary population modeling to infer viable post-supernova velocities and inspiral times. 
By populating binary tracers according to the star formation history of the host and kinematically evolving their post-supernova trajectories through the time-dependent galactic potential, we find that systems matching the observed offsets of the bursts require post-supernova systemic velocities of hundreds of kilometers per second. 
Marginalizing over uncertainties in the stellar mass--halo mass relation, we find that the second-born neutron star in the GRB~070809 and GRB~090515 progenitor systems received a natal kick of $\gtrsim 200~\mathrm{km\,s}^{-1}$ at the \GtrTwoHundredMarginalizedGRBOne and \GtrTwoHundredMarginalizedGRBTwo credible levels, respectively. 
Applying our analysis to the full catalog of localized short gamma-ray bursts will provide unique constraints on their progenitors and help unravel the selection effects inherent to observing transients that are highly offset with respect to their hosts.

\end{abstract}


\section{Introduction}

The association between cosmic transients and their galactic hosts holds clues to the evolution and formation of their progenitors. 
The locations of \acp{GRB} with respect to their host galaxies can be used as a key diagnostic to constrain their progenitor systems. 
For instance, long \acp{GRB} occur in star-forming host galaxies and have offsets that follow the exponential light profile of star-forming disks \citep{Bloom2002,Ramirez-Ruiz2002,Fruchter2006,Blanchard2016}, supporting their origin from young, massive-star progenitors. 
In contrast, short \acp{GRB} occur in both star-forming and quiescent galaxies~\citep{Prochaska2006,Fong2013,Berger2014}, and have offsets that typically exceed the effective radii of their hosts~\citep{Bloom2007,Berger2010,Fong2010,Fong2013a}. 
These offsets can be explained by a \ac{DNS} or neutron star--black hole origin~\citep{Eichler1989, Narayan1992} as a consequence of their broad delay-time distributions and their susceptibility to \ac{SN} kicks at formation. 
The direct link between \ac{DNS} mergers and short \acp{GRB} was established with the coincident observation of gravitational waves and a short \ac{GRB} from a \ac{DNS} merger with GW170817 and GRB~170817A~\citep{GW170817,GW170817_GRB,GW170817_MMA}. 

Since \ac{DNS} mergers are regarded as the dominant astrophysical mechanism for short \acp{GRB}, the population of short \acp{GRB} that are well localized and have robust host galaxy associations are a propitious route for constraining the properties of their \ac{DNS} progenitors~\citep[e.g.,][]{Bloom1999,Belczynski2006a,Bloom2007,Zheng2007,Troja2008,Kelley2010,Church2011,Fong2013a,Behroozi2014,GW170817_progenitor,Wiggins2018}. 
Of particular interest are the $\sim 15$--$20\%$ of short \acp{GRB} that have no coincident host galaxy to deep limits~\citep[e.g.,][]{Berger2010,Fong2013a,Berger2014,Tunnicliffe2014}; these hostless short \acp{GRB} likely migrated significant distances from their sites of formation.\footnote{Though often referred to as ``hostless'' in the literature, most short \acp{GRB} in this population have likely host associations, and therefore we instead use the nomenclature ``highly offset'' in this paper. }
GW170817 was only offset from its host by ${\approx 2~\mathrm{kpc}}$ in projection~\citep{Blanchard2017,Coulter2017}, whereas the projected physical separations for the highly offset population is $\approx 30$--$75~\mathrm{kpc}$ \citep{Berger2010}, which is typically $\gtrsim 5$ effective radii ($R_\mathrm{e}$) from their hosts~\citep{Fong2013a}. 
The need to explain these large offsets may challenge the paradigm that many \acp{DNS} receive low \ac{SN} kicks at birth~\citep{Podsiadlowski2004,vandenHeuvel2004,Schwab2010,Bray2016,Beniamini2016,Tauris2017}. 
However, to accurately constrain post-\ac{SN} barycentric velocities (i.e.\ systemic velocities) from short \ac{GRB} offsets, it is necessary to account for (i) the evolution and growth of host galaxies over cosmic time, (ii) the progenitor's motion in the host galaxy, and (iii) the interplay between \ac{SN} natal kicks and mass loss when determining the post-\ac{SN} motion. 

We present a forward modeling approach that follows the full kinematic evolution of short \ac{GRB} progenitors as their host galaxies evolve. 
Given that the theoretical and observational findings that dynamically formed \ac{DNS} systems do not contribute significantly to the overall merger rate \citep{Fong2019,Ye2020}, our study focuses on the predominant isolated binary evolution channel. 
Paired with population modeling, we constrain various aspects of the \ac{DNS} progenitors of two well-localized and highly offset short \acp{GRB}, GRB~070809 and GRB~090515, such as their post-\ac{SN} systemic velocities, inspiral times, and \ac{SN} natal kicks. 
We find that the \ac{DNS} mergers responsible for these events likely required substantial natal kicks at birth---in the case of GRB~090515, the lower limit on its \ac{SN} natal kick is consistent with the largest lower limits derived for Galactic \ac{DNS} natal kicks. 

In Section~\ref{sec:observations}, we briefly overview the observations and inferred properties of GRB~070809, GRB~090515, and their respective hosts. 
Section~\ref{sec:methods} covers the numerical methods used to model host galaxies, perform kinematic evolution of tracer particles, synthesize \ac{DNS} populations, and statistically determine constraints on progenitor parameters. 
In Section~\ref{sec:results}, we examine constraints on the kinematic evolution and \ac{DNS} progenitors from our analysis. 
Finally, in Section~\ref{sec:discussion} we summarize our main findings and compare our constraints to those from the Galactic \ac{DNS} population. 
Throughout the paper we use the \textit{Planck 2015} cosmological parameters: $H_0 = 68~\mathrm{km\,s}^{-1}\,\mathrm{Mpc}^{-1}$, $\Omega_\mathrm{m} = 0.31$, and $\Omega_{\Lambda} = 0.69$ \citep{PlanckCollaboration2016}.

\section{Host Galaxy Stellar Population Properties}\label{sec:observations}

We focus this study on two short \acp{GRB}, GRB~070809 and GRB~090515, which have no spatially coincident galaxy to $\gtrsim 26.2~\mathrm{mag}$ and $\gtrsim 26.5~\mathrm{mag}$, respectively \citep{Berger2010,Fong2013}. 
To determine likely host galaxy associations, previous studies have used probability of chance coincidence ($P_\mathrm{cc}$), which employs galaxy number counts to quantify the probability that a galaxy of a given apparent brightness is there by chance. 
Thus, a likely host galaxy for a given \ac{GRB} will have a low $P_\mathrm{cc}$ value. 
Using this metric, the most probable host galaxies for GRB~070809 and GRB~090515 are at offsets of $\approx 30$--$75~\mathrm{kpc}$ in projection (Table~\ref{table}). 
The likely hosts are both early-type galaxies, with no signs of ongoing star formation \citep{Rowlinson2010,Berger2010}.

To model their stellar populations, we use all available photometric observations of the host galaxies. 
For GRB~070809, we use ground-based $griK$-band observations \citep{Leibler2010}, as well as F606W and F160W photometry from the \textit{Hubble Space Telescope} \citep[\textit{HST};][]{Fong2013a}. 
For GRB~090515, we use ground-based $griJK$-bands, \citep{Leibler2010} and {\it HST}/F160W \citep{Fong2013a}. 
We also use the published spectroscopic redshifts of the hosts, the effective radii of the hosts, and projected physical offsets ($R_\mathrm{off}$) of the \ac{GRB} \citep{Fong2013a}. 

We fit for stellar population properties of the hosts with \texttt{Prospector}~\citep{Leja2016}, which uses the \texttt{Flexible Stellar Population Synthesis} code~\citep{Conroy2009} to build a stellar population model and determine its best-fitting properties based on available data. 
\texttt{Prospector} applies the \texttt{dynesty} nested sampling method~\citep{Speagle2020} to infer properties of the stellar population such as mass, age, dust extinction, \ac{SFH}, and metallicity. 

We perform fits with redshifts fixed to those of the \acp{GRB} (Table~\ref{table}) and set dust extinction $A_\mathrm{V}=0$~mag, as expected for quiescent galaxies. 
The maximum possible ages of the galaxies are determined by the age of the Universe at the respective redshifts of the short \acp{GRB} (GRB\,070809: $8.87~\mathrm{Gyr}$ and GRB\,090515: $9.40~\mathrm{Gyr}$). 
We include a prescription for the \ac{SFH}, parameterized by a delayed-$\tau$ function: $\textrm{\ac{SFH}} \propto t\exp({-t/\tau})$. 

We test a range of metallicities, $0.1$--$1 \mathrm{Z}_\odot$, leaving mass, stellar population age, and $\tau$ as free parameters. 
For both \acp{GRB}, the inferred galactic mass varies by $\lesssim 2\%$ across this metallicity range. 
On the other hand, metallicity is strongly degenerate with the stellar population age and causes the inferred age to increase with decreasing metallicity; at $0.1 \mathrm{Z}_\odot$, the recovered distribution on the stellar age pushes to extreme values and rails strongly against the age-of-Universe prior bound. 
However, we can quantify the quality of parameter fits at various metallicities by comparing their fully marginalized likelihoods (evidence). 
We find a strong preference for near-solar metallicities in both host galaxies; comparing to runs fixed at $\mathrm{Z}_\odot$ and $0.1 \mathrm{Z}_\odot$, we find log evidence ratios (log odds for different metallicities if they are considered equally likely a priori) of ${\ln[P(\mathrm{Z}_\odot)/P(0.1 \mathrm{Z}_\odot)]} = 228$ and $324$ for GRB~070809 and GRB~090515, respectively. 
As we discuss in Section~\ref{subsec:progenitor_constraints}, younger stellar ages (and therefore higher metallicities) also yield more conservative constraints on \ac{DNS} progenitor properties. 
For constructing our galactic hosts, we therefore fix the metallicities to $\mathrm{Z}_\odot$, as these runs best describe the observations and represent conservative constraints. 
Table~\ref{table} presents a summary of inferred properties from this work and the literature used in constructing our galaxy models.

\begin{deluxetable}{ l  c c}
\tablecaption{Observed and inferred parameters for short \acp{GRB} and their most probable host galaxies from \citet{Leibler2010}, \citet{Fong2013a}, and this work. 
68\% credible intervals are given for the observed offset ($R_\mathrm{off}^0$) as well as the inferred stellar mass ($M_\star^0$) and stellar age ($t_\star^0$). 
Superscripts $^0$ are used throughout the text to designate observed or inferred parameters that are used in our host galaxy modeling.  \label{table}}
\tablehead{
\colhead{} & \multicolumn{2}{c}{GRB} \\
\cline{2-3}
\colhead{\ac{GRB} and Host Properties} & 070809 & 090515
}
\startdata
Redshift $z^{0}$ & 0.47 & 0.40\\
Effective radius $R_\mathrm{e}^{0}$ [kpc] & $3.59$ & $4.24$ \\
Projected \ac{GRB} offset $R_\mathrm{off}^{0}$ [kpc] & $33.22\pm2.71$ & $75.03\pm0.15$ \\
Stellar mass $\log(M_\star^{0}/\mathrm{M}_{\odot})$ & $10.95^{+0.05}_{-0.01}$ & $10.87^{+0.03}_{-0.03}$ \\
Star formation rate $\dot{M}_\star^{0}$ [$\mathrm{M}_{\odot}\,\mathrm{yr}^{-1}$] & $\lesssim 0.1$ & $\lesssim 0.1$  \\
Stellar age $t_{\star}^{0}$ [Gyr] & $3.14^{+0.53}_{-0.16}$ & $5.49^{+0.60}_{-0.38}$ \\
Probability of chance coincidence $P_\mathrm{cc}$ & $0.03$ & $0.15$ \\
\enddata
\end{deluxetable}

\section{Modeling \ac{GRB} Progenitors}\label{sec:methods}

We use forward modeling to determine progenitor systems that match the observed redshift and projected offset of GRB~070809 and GRB~090515. 
We first construct a time-depedent potential for the host galaxy with three components for the stellar, gas, and \ac{DM} distributions, and populate the galaxy with tracer particles according to its gas-density profile and \ac{SFH}. 
We apply a post-\ac{SN} systemic velocity to the tracers, accounting for the motion in the galaxy before the \ac{SN}, and follow their evolution from birth until the redshift of the short \ac{GRB}. 
Finally, we identify tracers that match observations based on the offset constraints of the short \acp{GRB} to determine viable inspiral times and post-\ac{SN} systemic velocities, and combine these with distributions from population modeling to constrain properties of the progenitor systems.

\subsection{Host Galaxy Modeling}\label{subsec:galmodel}

The properties of \ac{GRB} hosts at the time of the explosion are often used as a proxy for the host galaxy at the time of progenitor formation. 
Though this may be a valid approximation for phenomena that occur shortly after their progenitor stars form, such as long \acp{GRB}, this assumption is inadequate for describing the evolution of systems that have broad delay-time distributions, as galaxies can evolve significantly since the time that the progenitor formed \citep[e.g.,][]{Kelley2010,Behroozi2014,Wiggins2018}. 
Therefore, galaxy evolution will play an important role when constraining aspects of short \ac{GRB} progenitors. 

To this end, we instead employ a time-dependent model for the galactic potentials of the \ac{GRB} hosts that accounts for galaxy growth along the \ac{SFMS}.
The \acp{SFH} of our model galaxies are calculated using a modified version of the procedure given in \cite{Speagle2014} along with their parameterization of the \ac{SFMS} of galaxies: $\dot{M}_\star = \dot{M}_\star(M_\star, t)$. 
Variances in the best-fit parameters from \cite{Speagle2014} are used to characterize the variance between galaxies along the \ac{SFMS}. 
Further, we assume that our model galaxies lie at a fixed percentile of the population relative to the median \ac{SFMS} relationship.  
Thus, a particular track in $\dot{M}_\star$--$M_\star$ space can be described as $\dot{M}_\star(M_\star(t), t, \sigma_{\dot{M}_\star}$), where $\sigma_{\dot{M}_\star}$ gives the distance of a particular galaxy from the median \ac{SFMS}, in units of standard deviations of the population. 
This is a reasonable approximation for galaxies that lie relatively close to the median of the \ac{SFMS} until star formation is shut off~\citep{Tacchella2016}. 
Both of the short \ac{GRB} host galaxies we examine are observed to be quenched---to have a low \ac{SFR} at the time of the short \ac{GRB} (see Table~\ref{table}). 
For this reason, the final parameter we introduce is a quenching time $\tau_q$ such that $\dot{M}_\star(t > \tau_q) = \dot{M}_{\star}^{0}$. 
We use the upper limits for $\dot{M}_{\star}^{0}$ as in Table \ref{table}, which produce a negligible change in the total mass of the galaxy after the quenching time. 

The short \ac{GRB} host observations and subsequent fits described in Section \ref{sec:observations} give the stellar mass of the galaxy ($M_\star^{0}$), and the median age of the stellar population ($t_\star^{0}$) at the time of the short \ac{GRB}.  
We construct a grid of \ac{SFH} tracks in $\dot{M}_\star$--$M_\star$ space, parametrized by $\sigma_{\dot{M}_\star}$ and $\tau_q$. 
From the \ac{SFH} grid we calculate the average stellar age and require that this matches the inferred value from observations of the host galaxy. 
This yields a set of valid parameters (a line if the age is assumed to be known exactly, or a band if uncertainty is included). 
To break this degeneracy we adopt the ansatz that the most probable $\sigma_{\dot{M}_\star}$ and $\tau_q$ are, respectively: zero, and halfway between the formation time corresponding to the typical stellar age and the time when the short \ac{GRB} occurred.\footnote{Both assumptions are motivated by the expectation value of these parameters: the former is the median observed \ac{SFMS} $\dot{M}_\star$, while the latter assumes a uniform probability between the available constraints on quenching time. } 

\begin{figure}[t!]\label{fig:kinematic}
\includegraphics[width=0.5\textwidth]{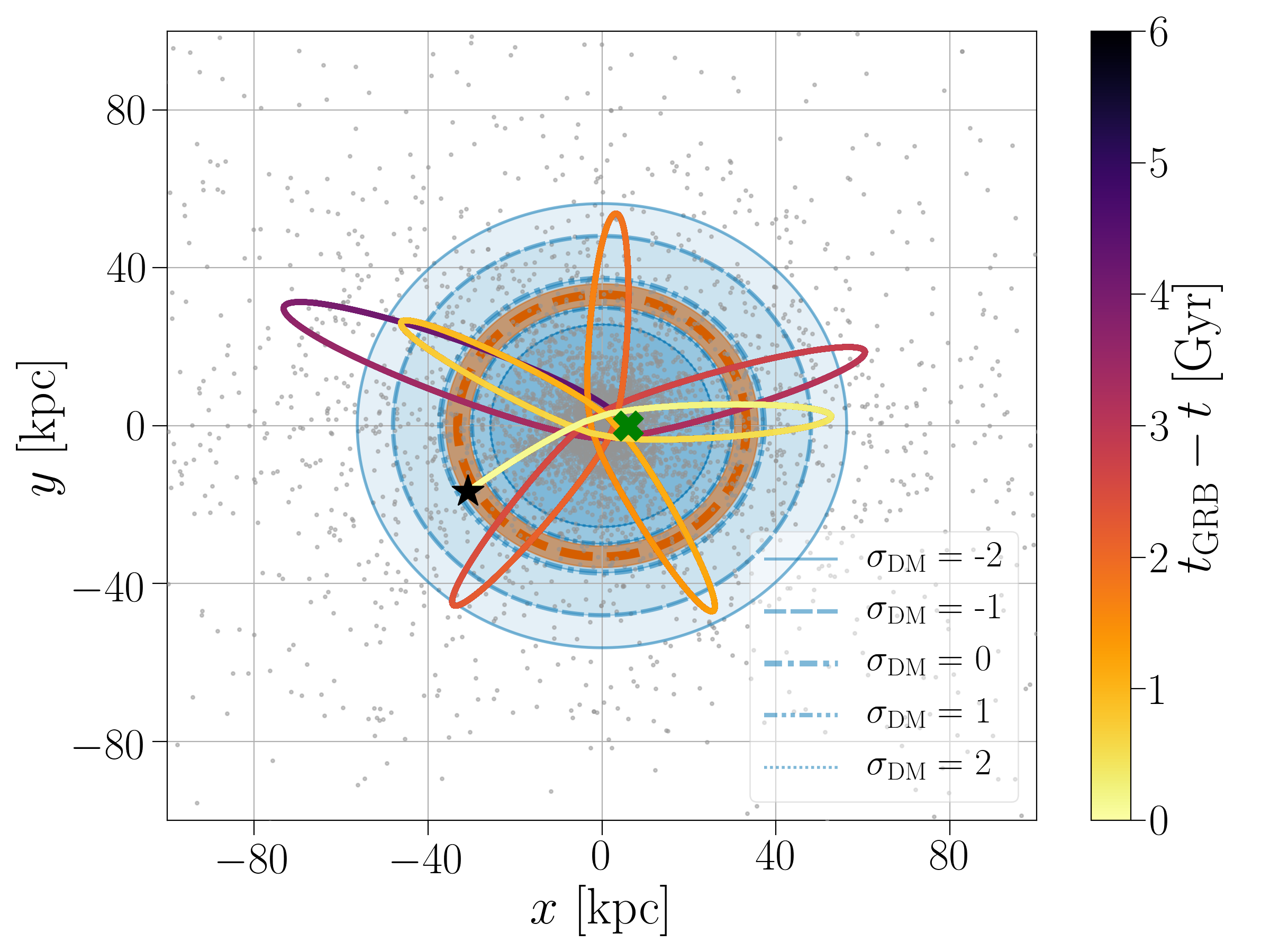}
\caption{Example kinematic evolution of a tracer particle that merges at the projected offset of GRB~070809. 
The green cross shows where the tracer was initiated (the location of the second \ac{SN}), and the black star is where the \ac{DNS} merges. 
The color of the trajectory denotes the passage of time (see color bar). 
The orange-dashed line and surrounding band mark $R_\mathrm{off}^{0} = 33.22\pm2.71$ kpc, the measured projected offset for GRB~070809. 
For five variations in the \ac{SMHM} relation, blue lines and shaded regions indicate the radii within which 95\% of tracers (weighted by the population prior) merge. 
Gray background points mark the locations where a subsample of simulated \ac{DNS} tracers merged. 
In this example, the \ac{DNS} system migrated a net distance of \DistanceTraveled between formation and merger, and has a post-\ac{SN} systemic velocity of \PostSNSystemic. 
}
\end{figure}

Details regarding the construction of our three-dimensional, time-dependent, multicomponent galaxy models can be found in Appendix~\ref{app:hostgal}. 
To summarize, the galactic potential is completely determined by the scale radius of the star-forming disk and component mass at each time $t$, which themselves are inferred using only the limited properties known for the host galaxies: $M_{\star}^{0}$, $z^{0}$, $R_\mathrm{e}^{0}$, $\dot{M}_\star^{0}$, and $t_{\star}^{0}$. 
We construct five models for each host galaxy with $\{0,\pm1\,\sigma, \pm2\,\sigma\}$ deviations from the median \ac{SMHM} relation from \cite{Moster2013} to investigate the sensitivity of our results to assumptions about the \ac{DM} halo mass. 
We then use the \texttt{galpy} package~\citep{galpy} to construct the galactic potentials and to create interpolation models of the potentials to speed up the kinematic integration of tracer particles, as described in the following section.

\subsection{Kinematic Evolution}

We kinematically evolve $10^6$ tracer particles (representing \ac{DNS} systems) in each host galaxy model. 
The time at which tracers are initialized in the host (when the \ac{DNS} systems are born) is determined by the \ac{SFH} of the model galaxy. 
Details of how tracer particles are populated can be found in Appendix~\ref{app:tracers}. 

The orbit of each tracer is integrated in the time-dependent galactic potential $\Phi(t)$ using \texttt{galpy}~\citep{galpy} until the physical time that the \ac{GRB} occurred. 
Our final results are all marginalized over random lines of sight. 
The kinematic evolution of a tracer particle whose final projected offset matches $R_\mathrm{off}^{0}$ for GRB~070809 is shown in Figure~\ref{fig:kinematic}. 

This procedure results in a distribution of tracer particle offsets as a function of time for each \ac{GRB} host galaxy. 
However, the method described so far is agnostic to the typical inspiral times and systemic velocities that result from an astrophysical population of \ac{DNS} progenitors. 
To properly describe the radial offset distribution of short \acp{GRB}, we convolve the systemic velocities and evolution times of our tracer particles with those anticipated for an astrophysical \ac{DNS} population, which we describe in the following sections. 
This methodology allows one to easily modify the input population models, which come with their own inherent uncertainties.

\subsection{Population Modeling}\label{subsec:popmodel}

We use the population synthesis code \texttt{COSMIC}\footnote{\href{https://cosmic-popsynth.github.io/}{cosmic-popsynth.github.io} (Version 3.2)}~\citep{Breivik2020} to model \ac{DNS} populations at the birth of the second neutron star. 
\texttt{COSMIC} is based upon a modified version of \texttt{BSE}~\citep{Hurley2002}, with updates to include state-of-the-art prescriptions for mass loss in O and B stars \citep{Vink2001}, metallicity dependence in the evolution of Wolf--Rayet stars \citep{Vink2005}, new prescriptions for fallback and post-\ac{SN} remnant masses \citep{Fryer2012}, variable prescriptions for the common envelope $\lambda$ parameter \citep{Claeys2014}, as well as procedures for implementing electron-capture \acp{SN} \citep{Podsiadlowski2004} and ultra-stripped \acp{SN} \citep[][]{Tauris2013,Tauris2015}. 

We model a population of \ac{DNS} systems at $\mathrm{Z}_\odot$, consistent with our metallicity assumption for the host galaxy models, and extract their systemic velocities and inspiral times immediately following the second \ac{SN}. 
Though metallicity strongly affects \ac{DNS} merger rates, properties of merging \acp{DNS} vary only slightly with metallicity \citep{Dominik2012,Giacobbo2018a,Chruslinska2019,Neijssel2019}, and thus our choice of metallicity should have a minor impact on our results; we discuss the sensitivity of our results to the assumed metallicity of our stellar populations in Section \ref{subsec:progenitor_constraints}. 
These astrophysically motivated distributions can then be convolved with the associated variables in our kinematic modeling to attain a relative weighting for each tracer.

\subsection{Identifying Short \ac{GRB} Analogs}\label{subsec:bayesian}

In our analysis, we have three general sets of parameters. 
The observed parameters, $\vec{\Theta}_\mathrm{obs} = \{z^{0}, R_\mathrm{off}^{0}, \sigma_{R_\mathrm{off}^{0}}\}$, act to constrain the viable inspiral times and offsets from our kinematic analysis that match the redshift and offset of a given \ac{GRB}. 
The kinematic parameters, $\vec{\Theta}_\mathrm{kin} = \{V_\mathrm{sys},\,R_\mathrm{birth},\,z_\mathrm{birth}\}$, are aspects of our kinematic modeling, and though $R_\mathrm{birth}$ and $z_\mathrm{birth}$ are dependent on the galaxy modeling, $\vec{\Theta}_\mathrm{kin}$ are agnostic to particulars of binary evolution and \ac{DNS} formation. 
Finally, the population parameters, $\vec{\Theta}_\mathrm{pop} = \{V_\mathrm{k}, A_\mathrm{pre}, M_\mathrm{pre}, M_\mathrm{post}, M_{2}\}$, which are the magnitude of the second \ac{SN} natal kick, the pre-second \ac{SN} semi-major axis, the pre-second \ac{SN} mass of the exploding star, the post-second \ac{SN} mass of the exploding star, and the companion mass at the second \ac{SN}, respectively, are the variables that map the properties of the binary at the second \ac{SN} to the systemic velocity and inspiral time following the second \ac{SN}. 

We ultimately wish to determine constraints on the \ac{DNS} progenitor properties from the observed properties of the \ac{GRB},  $p(\vec{\Theta}_\mathrm{pop} | \vec{\Theta}_\mathrm{obs})$. 
As a first step, we examine the constraints on our kinematic model parameters, $\vec{\Theta}_\mathrm{kin}$, given the observed parameters, $\vec{\Theta}_\mathrm{obs}$: 
\begin{equation}\label{eq:kin_posterior}
    p(\vec{\Theta}_\mathrm{kin} | \vec{\Theta}_\mathrm{obs}) \propto \int w_\mathrm{obs} \times w_\mathrm{kin} \times w_\mathrm{gal} \times  w_\mathrm{pop}\,\mathrm{d}R_\mathrm{off}\,\mathrm{d}z. 
\end{equation}
where $w_\mathrm{obs}$ is the observational weighting of the likelihood, $w_\mathrm{kin}$ is the result of our kinematic modeling, $w_\mathrm{gal}$ is the weighting on birth location and inspiral time from the galaxy model, and $w_\mathrm{pop}$ is the weighting of inspiral times and systemic velocities from \ac{DNS} population synthesis (see Appendix~\ref{app:bayesian} for details). 
From here, the posterior distribution on population parameters follows from again invoking Bayes' theorem on $w_\mathrm{pop}$, which leads to 
\begin{equation}\label{eq:pop_posterior}
\begin{aligned}
    p(\vec{\Theta}_\mathrm{pop} | &\vec{\Theta}_\mathrm{obs}) 
    \propto \int w_\mathrm{obs} w_\mathrm{kin} w_\mathrm{gal} \\
    &\times p(V_\mathrm{sys}, t_\mathrm{insp} | \vec{\Theta}_\mathrm{pop}) p(\vec{\Theta}_\mathrm{pop}) \,\mathrm{d}R_\mathrm{off} \,\mathrm{d}z \,\mathrm{d}V_\mathrm{sys} \,\mathrm{d}t_\mathrm{insp}. 
\end{aligned}
\end{equation}
A full derivation for the posterior distributions $p(\vec{\Theta}_\mathrm{kin} | \vec{\Theta}_\mathrm{obs})$ and $p(\vec{\Theta}_\mathrm{pop} | \vec{\Theta}_\mathrm{obs})$ can be found in Appendix \ref{app:bayesian}.

\begin{figure*}[t!]\label{fig:Rproj_offset}
\includegraphics[width=0.95\textwidth]{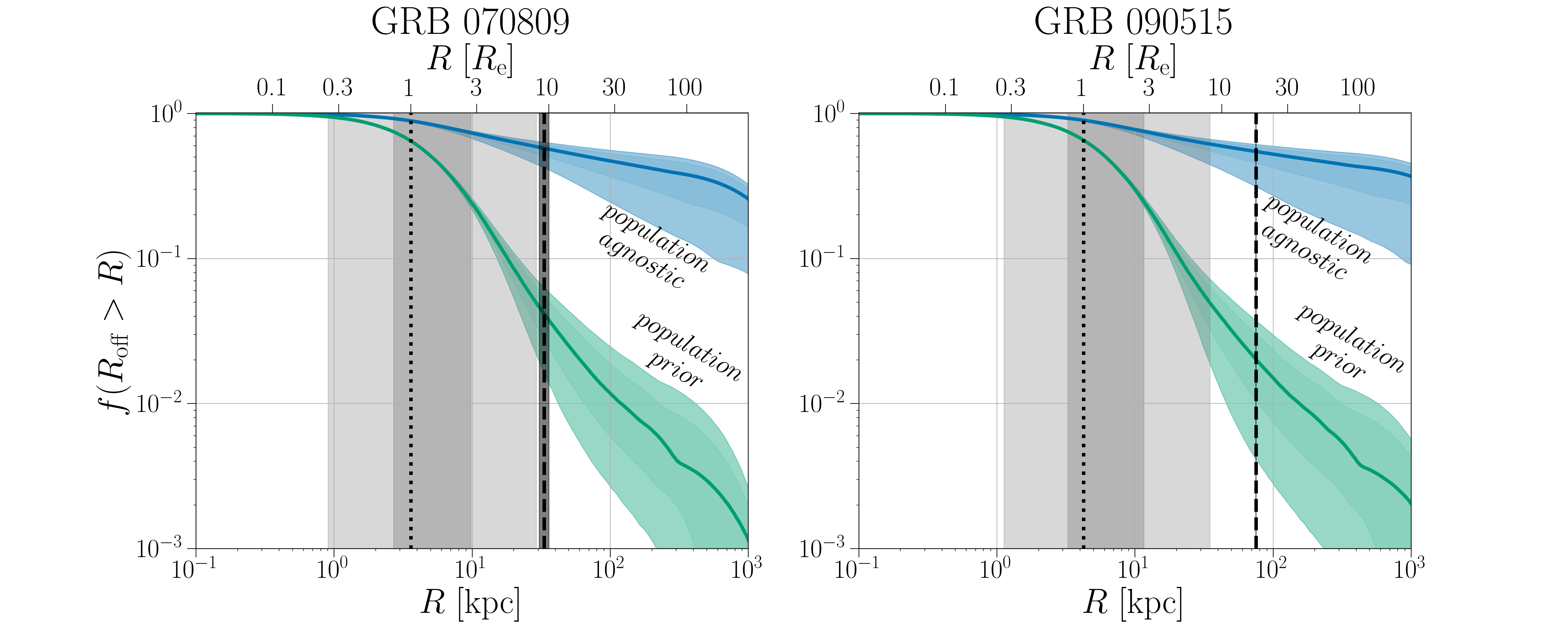}
\caption{Cumulative distribution function of projected offsets in each galaxy model at the time of the \ac{GRB}. 
In blue, the tracers used to construct the cumulative distributions are weighted uniformly in systemic velocity, and inspiral times are drawn according to the galactic \ac{SFH}. 
In green, the tracers are also weighted by the population modeling prior $w_\mathrm{pop}$, as described in Section \ref{subsec:bayesian}. 
The solid line shows the cumulative distributions for the median of the \ac{SMHM} relation, and the dark and light colored bands show the offset for models where the \ac{DM} halo masses are $1\,\sigma$ and $2\,\sigma$ above and below the median. 
Assumptions about the \ac{SMHM} relation have little impact at low offsets, though lower-mass halos lead to larger tails at the high end of the offset distribution. 
Gray bands show the $50\%$ and $90\%$ credible regions of the projected offset distribution, weighted by the population prior, for the $\sigma_\mathrm{DM} = 0$ model. 
The dotted black line marks the effective radius of the host galaxy at the time of the \ac{GRB}, and the dashed black line and dark gray shading marks the location and uncertainty of the \ac{GRB} projected offset. 
}
\end{figure*}

\section{Results}\label{sec:results}

By combining galaxy modeling, kinematic evolution of tracer particles, and \ac{DNS} population synthesis we can thoroughly examine the origins of short \acp{GRB} that occur at large offsets from their hosts. 
In addition to yielding constraints on \ac{DNS} progenitor properties, we can see how anomalous certain short \ac{GRB} systems are, and gain a better grasp on the selection effects inherent to such observations. 
We first discuss the implications of our results on offset distributions, both at the time of the \acp{GRB} and throughout cosmic time. 
We then place constraints on \ac{DNS} progenitor parameters, such as \ac{SN} kicks and mass loss, that are consistent with the observations of GRB~070809 and GRB~090515.

\begin{figure*}[t!]\label{fig:weighted_offset}
\includegraphics[width=0.98\textwidth]{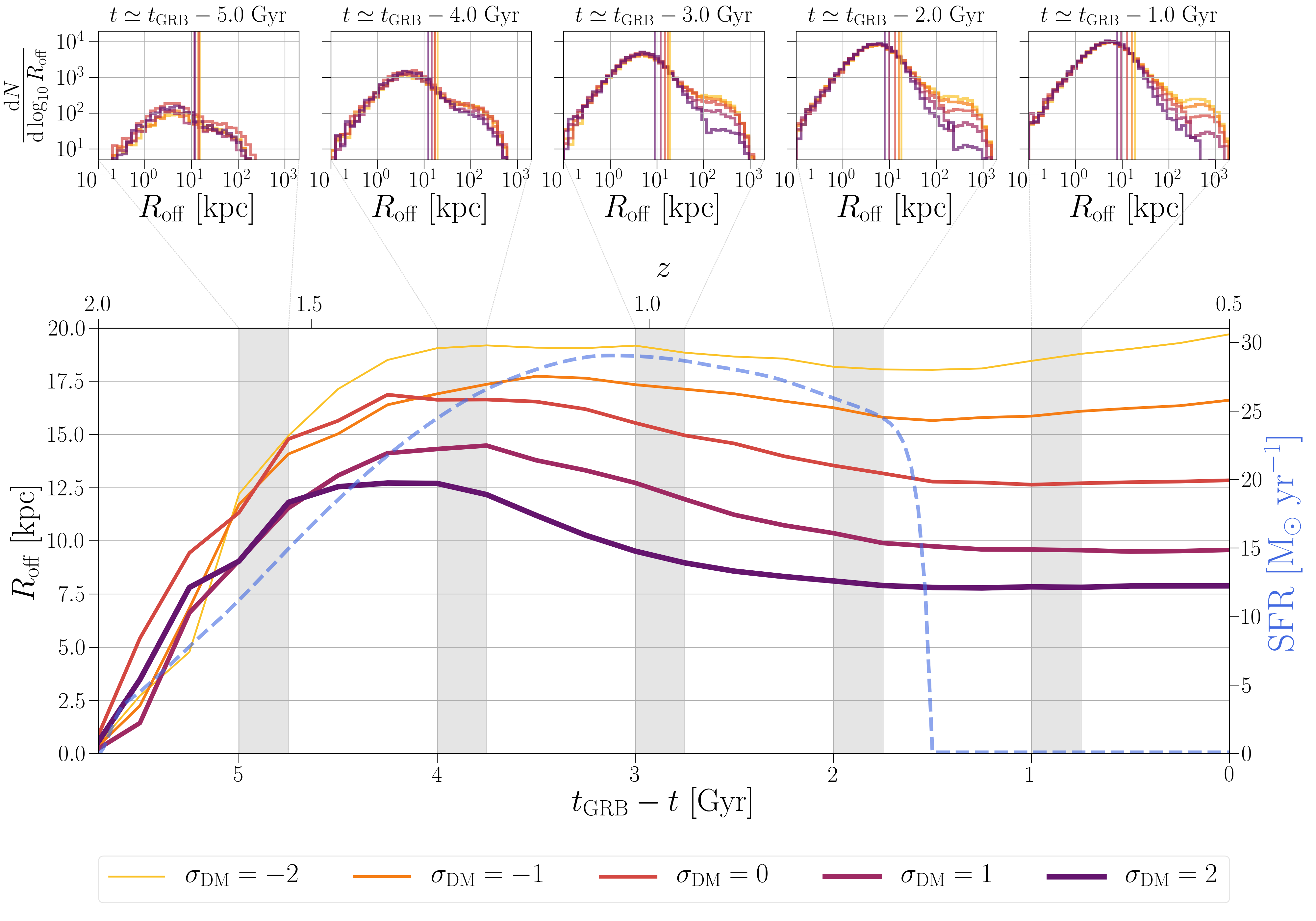}
\caption{Weighted projected offset distribution of tracer particles in the GRB~070809 host over cosmic time. 
Bottom panel: colored lines represent the mean tracer offsets in five models with varying deviations from the median \ac{SMHM} relation. 
At each point in time, only tracers that are injected into the model \textit{before} that time are included in the weighted average. 
The blue-dashed line denotes the \ac{SFR} of the host for reference, analogous to the galaxy model in Figure~\ref{fig:galaxy_model}. 
Gray bands and upper insets show the offset distribution at different slices of time in the galaxy's history, with vertical lines marking the mean of the distribution at that time slice. 
The height of the histograms also demonstrates the total number of kinematic tracers from the full  sample that are evolving in the galaxy at each point in time, a proxy for the relative number of \ac{GRB} progenitor systems at each point in time. 
}
\end{figure*}

\subsection{Coupling Galaxy Evolution with Progenitor Kinematic Modeling}

Both GRB~070809 and GRB~090515 are highly offset from their host galaxies, with projected offsets of ${\simeq 33~\mathrm{kpc}}$ (${9.25\,R_\mathrm{e}}$) and ${\simeq 75~\mathrm{kpc}}$ (${17.70\,R_\mathrm{e}}$), respectively. 
Figure~\ref{fig:Rproj_offset} shows the weighted projected offset distribution for tracer particles in the GRB~070809 and GRB~090515 host galaxy models, at the time of each \ac{GRB}. 
Here we do not constrain the offset of the \ac{DNS} tracers so we can examine all \ac{GRB} candidates from the host models rather than just those that match the observed offset of the \ac{GRB}. 

The projected offset of GRB~070809 falls slightly outside of the symmetric $90\%$ credible region, with \PercentileGRBOne of \ac{DNS} mergers from its host merging at larger offsets in our $\sigma_\mathrm{DM} = 0$ model. 
For the host of GRB~090515, the projected offset distribution pushes to slightly larger values; \PercentileFiftyGRBTwo of \ac{DNS} mergers occur at $\gtrsim 50~\mathrm{kpc}$ for the host of GRB~090515 compared to \PercentileFiftyGRBOne of \ac{DNS} mergers for the host of GRB~070809. 
This is due to the galaxy's quiescent phase occurring at an earlier time, allowing ejected tracers from the time of peak star formation to diffuse outward for longer and achieve more extreme offsets. 
GRB~090515's offset at the \PercentileGRBTwo percentile in our $\sigma_\mathrm{DM} = 0$ model, indicating that it is extreme, but not inexplicable, especially when considering that it has one of the highest offsets in the well-localized short \ac{GRB} population. 
We find that \FiveReffApprox (\TenReffApprox) of mergers occur at $\gtrsim 5\,R_\mathrm{e}$ ($\gtrsim 10\,R_\mathrm{e}$) in these particular hosts (see Figure~\ref{fig:Rproj_offset}). 

It is also informative to examine how the offset distribution evolves as a function of time. 
In the bottom panel of Figure~\ref{fig:weighted_offset}, we plot the evolution of the projected offset distribution over cosmic time for the host of GRB~070809. 
The value at each point in time is the mean of the weighted projected offset for all tracers that are injected into the galaxy before that time; as the galaxy grows and the \ac{SFR} increases, more tracers are populated into the model and incorporated into the weighted mean. 
At early times, few tracers exist in the galaxy (\NTracersFiveGyr of tracers are injected earlier than $t_\mathrm{GRB} - 5~\mathrm{Gyr}$) and the offset distribution is consistent across differing assumptions about the \ac{DM} halo mass. 
However, as can be seen in the top panels of Figure~\ref{fig:weighted_offset}, as time progresses the distributions begin to diverge at large offsets since lower-mass halos allow for tracers to migrate farther from their hosts, in some cases $\gtrsim 1~\mathrm{Mpc}$. 

The mean projected offset of tracers is larger at early times (${\sim 4~\mathrm{Gyr}}$ before the short \ac{GRB}) compared to late times. 
This is because the total mass of the host is smaller and gravitational potential shallower, allowing for tracers to explore a larger volume of their host's outskirts early on. 
However, there are relatively few tracers evolving at these early times since the injection of tracers is proportional to the \ac{SFR}. 
As more tracers are injected at later times, the mean offset steadily decreases since the galaxy's mass and potential well grow---though these tracers typically have larger pre-\ac{SN} galactic velocities, they are embedded in a deeper gravitational potential and do not reach the offsets of their predecessors. 
Once star formation shuts off, at $\sim\,t_\mathrm{GRB} - \TQuenchGRBOne$ for the host of GRB~070809, the decline in the offset distribution ceases. 
Since few new stars are being born at late times after quenching has completed (\MassAfterQuenching of the stellar mass budget), the offset distribution has a slow rise due to unbound tracers diffusing away from their hosts.

\begin{figure}[b!]\label{fig:kinematic_corner}
\includegraphics[width=0.46\textwidth]{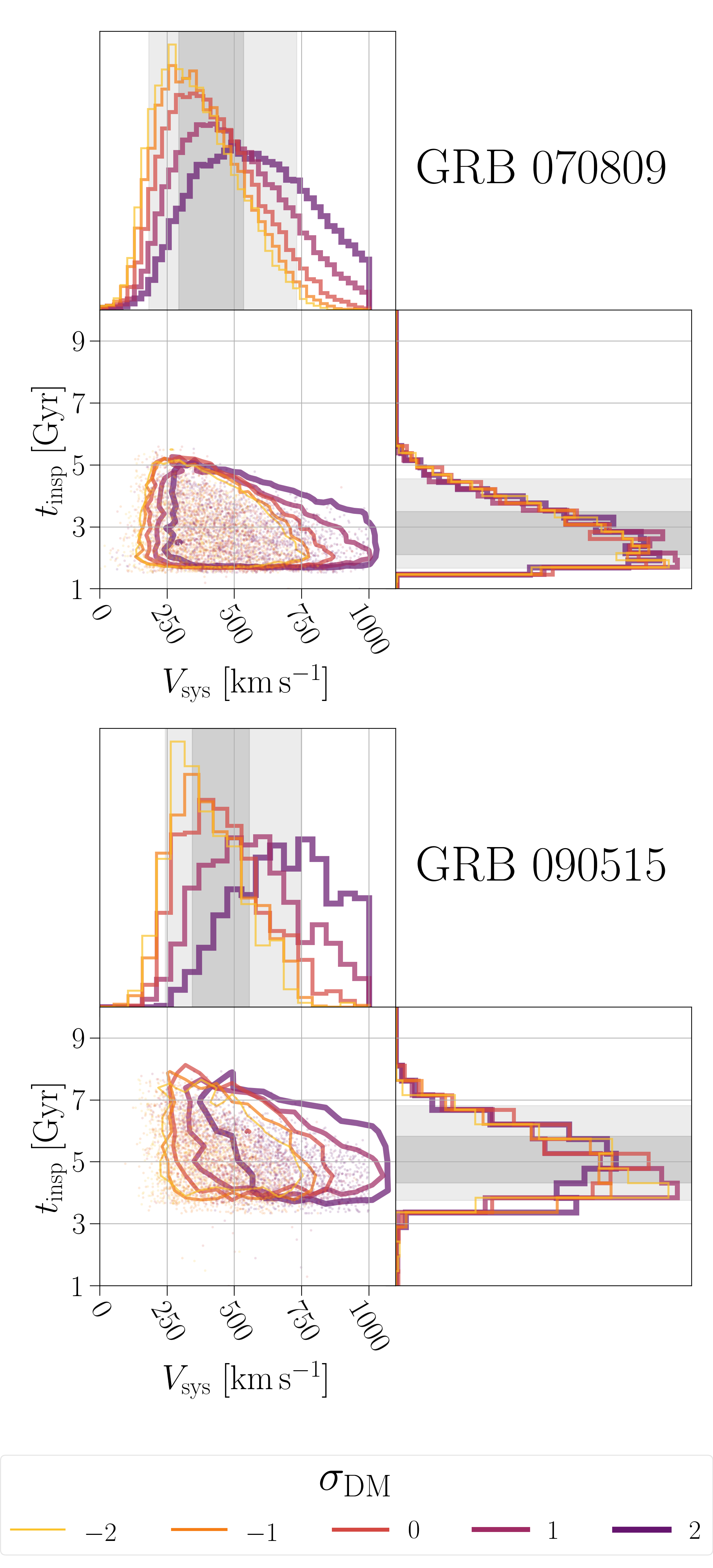}
\caption{Posterior distribution on the post-\ac{SN} systemic velocity $V_\mathrm{sys}$ and inspiral time $t_\mathrm{insp}$. 
The population prior, $w_\mathrm{pop}$, is set to unity to examine the constraints solely due to the observed projected offset of the \ac{GRB}. 
Colored lines represent different deviations from the median \ac{SMHM} relation. 
$90\%$ credible regions are shown in the joint posterior for each \ac{SMHM} model, with colored points showing samples from these distributions. 
Dark and light gray bands on the marginalized posteriors show the $50\%$ and $90\%$ credible intervals in the $\sigma_{\rm DM}=0$ model. 
}
\end{figure}

\begin{figure*}[t!]\label{fig:Vkick_constrains}
\includegraphics[width=0.95\textwidth]{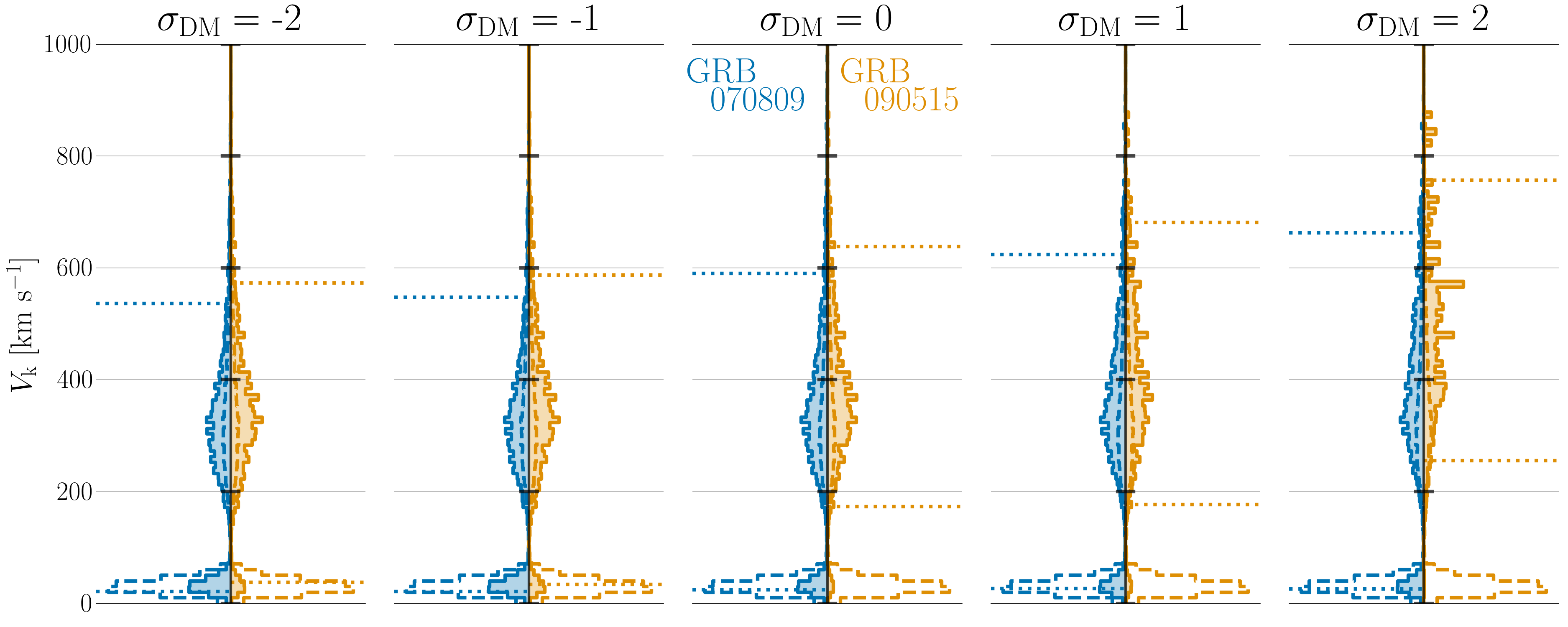}
\caption{Posterior distributions on natal kick magnitudes using five \ac{SMHM} relation realizations for the hosts of GRB~070809 (left, blue) and GRB~090515 (right, orange). 
Dashed lines show the prior that comes from population modeling, $p(\vec{\Theta}_\mathrm{pop})$, and solid lines/filled histograms show the posterior distribution, $p(\vec{\Theta}_\mathrm{pop} | \vec{\Theta}_\mathrm{obs})$. 
Dotted lines mark the lower and upper bounds of the symmetric $90\%$ credible interval. 
}
\end{figure*}

In Figure~\ref{fig:kinematic_corner} we show the joint posterior distribution on systemic velocities and inspiral times for various assumptions about the \ac{SMHM} relation. 
We set $w_\mathrm{pop}$ to unity to explore how the observed offset alone informs the viable inspiral times and systemic velocities. 
Systemic velocities are pushed to larger values as the \ac{DM} halo mass increases; for GRB~070809 (GRB~090515), $90\%$ of systemic velocities are above \VsysNinteyLowDMGRBOne (\VsysNinteyLowDMGRBTwo) for $\sigma_{\rm DM}=-2$ whereas $90\%$ of systemic velocities are above \VsysNinteyHighDMGRBOne (\VsysNinteyHighDMGRBTwo) for $\sigma_{\rm DM}=2$. 
This indicates that even for low assumed \ac{DM} halo masses, both \acp{GRB} require significant post-\ac{SN} systemic velocities in order to explain their observed projected offset. 

The inspiral times are less sensitive to our assumptions about the \ac{SMHM} relation, though for GRB~090515 longer inspiral times are somewhat preferred as we push to higher \ac{DM} halo masses. 
For the $\sigma_\mathrm{DM} = 0$ model, we find median inspiral times of \TinspMedGRBOne and \TinspMedGRBTwo for GRB~070809 and GRB~090515, respectively. 
These values are slightly lower than the typical stellar age in the two host galaxies (see Table~\ref{table}), indicating a preference for shorter inspiral times solely from the kinematic modeling. 
The insensitivity of the inspiral time on the \ac{DM} halo mass is due to these particular hosts being old, quiescent galaxies that formed the bulk of their stellar population $\gtrsim 1~\mathrm{Gyr}$ prior to the short \ac{GRB}, and the \ac{DNS} progenitors of the \ac{GRB} having ample time to explore the galactic environment prior to merging. 

These posterior distributions are attained by assuming a flat prior on systemic velocities, and a prior on inspiral times that corresponds to the \ac{SFH} of the host galaxy. 
However, the astrophysical distribution of systemic velocities and inspiral times expected for \ac{DNS} systems will add another term to the prior and influence the recovered distributions of these parameters, which we examine in the following section.

\subsection{Short \ac{GRB} Progenitor Constraints}\label{subsec:progenitor_constraints}

Binary stellar evolution is a poorly understood process, particularly for massive stars. 
The complicated mapping between stellar initial conditions and the birth properties of compact remnants can depend on many uncertain physical processes, such as two \acp{SN}, multiple phases of mass transfer, mass loss, stellar rotation, and tidal interactions (see \citealt{DeMink2015} and references therein for a review). 
However, \ac{DNS} properties at formation are dependent on a relatively small number of parameters of the binary system at the second \ac{SN}, namely the magnitude (and direction) of the \ac{SN} natal kick, the pre-\ac{SN} mass of the exploding star, the compact remnant masses, and the pre-\ac{SN} orbital separation and eccentricity \citep{Andrews2019}. 

Figure~\ref{fig:Vkick_constrains} shows the posterior distributions for the natal kick magnitudes of the second \ac{SN}. 
We marginalize over the other pre-\ac{SN} parameters that impact the post-\ac{SN} systemic velocity and inspiral time. 
The prior distribution used to draw natal kicks in our population model is shown with dashed lines, and consists of two Maxwellian distributions: the broad distribution at higher velocities represents systems that underwent a standard iron core-collapse \ac{SN}, whereas the narrower distribution at lower velocities represents systems that either underwent an electron-capture \ac{SN} or an ultra-stripped \ac{SN} \citep{Vigna-Gomez2018,Zevin2019b}.

With the solid lines, we show the posterior distribution on natal kick magnitudes for both GRB~070809 and GRB~090515, across our five realizations of the \ac{SMHM} relation for their host galaxies. 
To marginalize over our uncertainty in $\sigma_{\mathrm{DM}}$, we weight samples from each model according to their deviation from the \ac{SMHM} relation: 
\begin{equation}
    w_{i} = \frac{\mathcal{N}(\sigma_{\mathrm{DM},i};\,0,1)}{\sum_{i}\mathcal{N}(\sigma_{\mathrm{DM},i};\,0,1)}
\end{equation}
where $\mathcal{N}(\sigma_{\mathrm{DM},i};0,1)$ is a zero-mean Gaussian distribution with a standard deviation of unity evaluated at the deviation from the mean \ac{SMHM} relation in model $i$. 
For GRB~070809, large natal kicks from iron core-collapse \acp{SN} are slightly preferred compared to the prior. 
We find that \GtrTwoHundredMarginalizedGRBOne of systems matching the offset of GRB~070809 received natal kicks above $200~\mathrm{km\,s}^{-1}$, though as can be seen in Figure~\ref{fig:Vkick_constrains} low natal kicks of ${\lesssim 50~\mathrm{km\,s}^{-1}}$ are still consistent with the observed offset. 
On the other hand, after marginalizing over our $\sigma_{\mathrm{DM}}$ models, we find $\gtrsim\,\GtrTwoHundredMarginalizedGRBTwo$ of systems matching GRB~090515 have natal kicks of ${> 200~\mathrm{km\,s}^{-1}}$ and that low natal kicks  of ${\lesssim 50~\mathrm{km\,s}^{-1}}$ are strongly disfavored across all models, demonstrating that this system most likely received a substantial natal kick at birth to explain its observed offset. 
This is particularly apparent for larger halo masses; for \ac{DM} halos that are $2\,\sigma$ above the median \ac{SMHM} relation, less than \LowKickPercentageTwoSigamGRBTwo of \ac{DNS} tracers matching the observed offset have natal kick velocities of ${50~\mathrm{km\,s}^{-1}}$ or less. 

\begin{figure}[t!]\label{fig:param_constraints}
\includegraphics[width=0.48\textwidth]{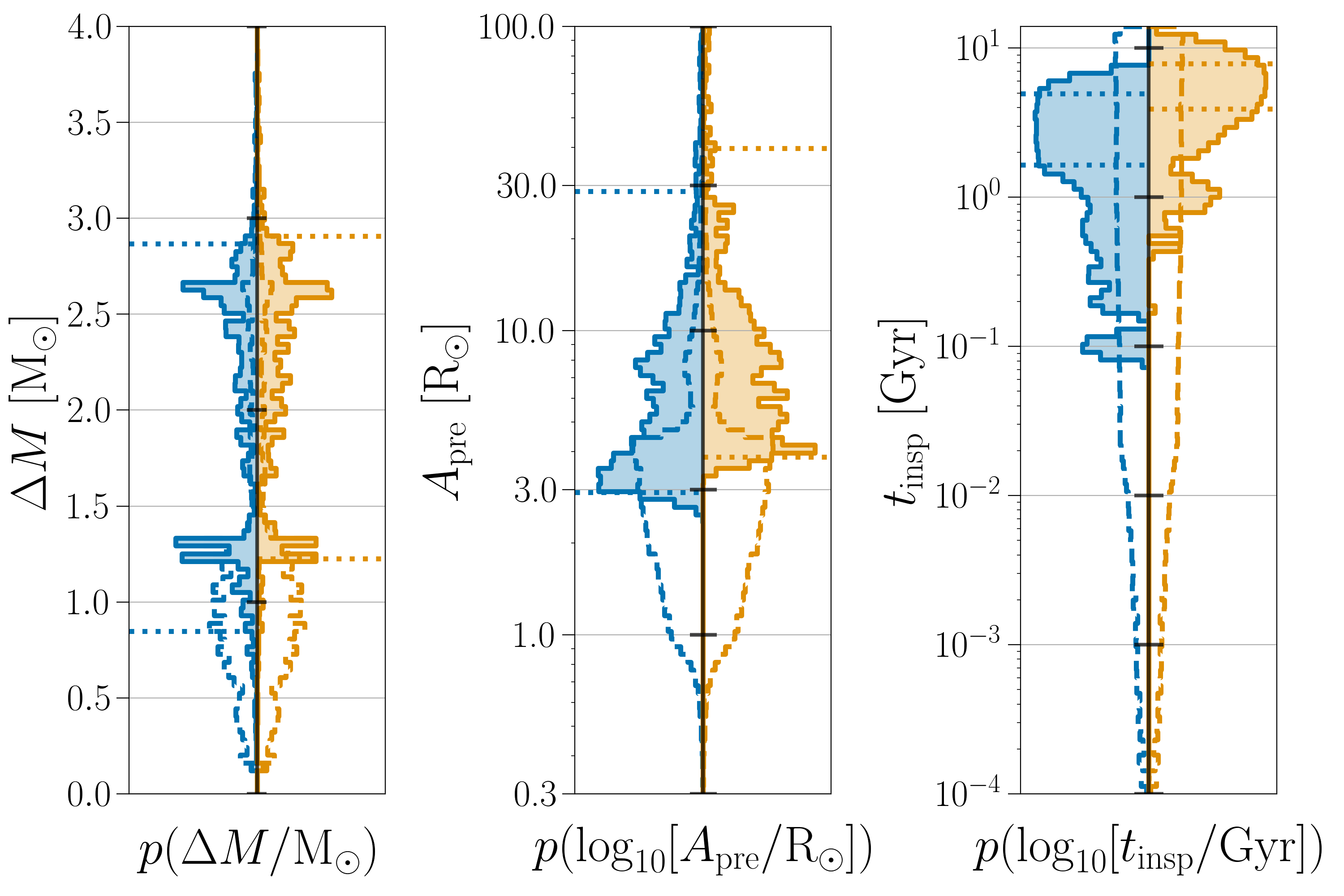}
\caption{Same as Figure~\ref{fig:Vkick_constrains}, but for various parameters describing the compact binary progenitors at the time of the second \ac{SN}. 
For the weighted $t_\mathrm{insp}$ posterior distributions, normalized bin heights below $10^{-8}$ are set as the zero point of the histogram height scale. 
Galaxy realizations assume the median value for the \ac{SMHM} relation ($\sigma_\mathrm{DM} = 0$). 
We see a slight trend to larger amounts of mass loss when we assume more massive \ac{DM} halos, however, the pre-\ac{SN} separation and inspiral time are relatively insensitive to deviations from the median \ac{SMHM} relation. 
}
\end{figure}

The post-\ac{SN} systemic velocity and \ac{DNS} inspiral time are also affected by the mass loss in the \ac{SN} explosion and the pre-\ac{SN} orbital separation. 
Assuming symmetric mass loss in the frame of the exploding star, conservation of momentum leads to a barycentric kick to the binary~\citep{Blaauw1961}, which scales as $\Delta M_\mathrm{SN} / \sqrt{A_\mathrm{pre}}$ to leading order. 
The combination of this mass-loss kick and the natal kick determine the post-\ac{SN} systemic velocity and orbital properties (and thereby the \ac{DNS} inspiral time). 
In Figure~\ref{fig:param_constraints}, we show the marginalized posterior distributions for mass loss and pre-\ac{SN} separation, as well as the inspiral times they map to. 

Though inspiral times for \acp{DNS} can be as low as ${10^4~\mathrm{yr}}$, we see that short inspiral times are disfavored (particularly for GRB~090515), since the bulk of star formation in the host galaxies occurred before the galaxies were quenched at $t_{\rm GRB} - t \simeq \TQuenchGRBOne$ and $t_{\rm GRB} - t \simeq \TQuenchGRBTwo$ for GRB~070809 and GRB~090515, respectively. 
The preference for larger inspiral times shows corresponding effects in the posterior distributions for mass loss and pre-\ac{SN} separation. 
Tight pre-\ac{SN} orbital separations of $\lesssim 3\,\mathrm{R}_\odot$ are disfavored for both GRB~070809 and GRB~090515, since these correspond to \acp{DNS} being born after galactic star formation has quenched. 
Additionally, since relatively large post-\ac{SN} systemic velocities are required for \acp{DNS} to migrate to the observed offset of merger, the larger orbital separations must be mitigated by larger amounts of mass loss (left panel of Figure~\ref{fig:param_constraints}) as well as larger natal kicks.

Our choice of metallicity acts as a conservative lower limit for the natal kicks recovered by our analysis. 
We assume stellar populations at solar metallicity, both for modeling the evolution of the host galaxy and for generating \ac{DNS} populations. 
As discussed in Section \ref{sec:observations}, metallicity is largely degenerate with the age of the stellar population inferred from observations, such that assuming a lower metallicity causes the stellar population age to increase. 
This will result in longer inspiral times, and therefore larger pre-\ac{SN} orbital separations, to match the observed offset and merger time of the short \ac{GRB}. 
Though lower metallicity stars may lead to more mass loss in the \ac{SN} due to weaker stellar winds earlier in the progenitor's evolution, larger natal kicks will still be needed to compensate for the increase in pre-\ac{SN} orbital separation since at larger separations the mass loss in the \ac{SN} will have a lesser impact on the post-\ac{SN} systemic velocity. 
Exploring deeper how variations in metallicity impact our inference is reserved for future work.

\section{Discussion and Conclusions}\label{sec:discussion}

The galactic host associations of short \acp{GRB} embed information about their compact binary progenitors. 
In this study, we leverage spectroscopic and photometric observations of short \acp{GRB} and their hosts to construct empirically motivated host galaxy models and examine the kinematic evolution of \ac{DNS} progenitors over cosmic time. 
In doing so, we place \ac{GRB}-specific constraints on the typical velocities and inspiral times of their \ac{DNS} progenitors, and for the first time, pair the kinematic analysis with population modeling to infer properties of short \ac{GRB} progenitors at the time of \ac{DNS} formation.

\subsection{Comparison to Galactic \ac{DNS} Properties}

Much of our observational knowledge about \ac{DNS} mergers comes from the small, but growing catalog of \ac{DNS} systems observed in the Milky Way. 
As of now, 19 Galactic \acp{DNS} have been discovered, with the majority residing in the Galactic field~(see \citealt{Tauris2017, Andrews2019a}, and references therein).
The present-day orbital properties and neutron star masses offer insights into the explosion mechanisms that form compact objects~\citep{Wex2000,Podsiadlowski2004,Schwab2010,Wong2010,Andrews2015,Tauris2017,Beniamini2016,Bray2016,Vigna-Gomez2018,Andrews2019a}. 
Pairing this information with the present-day motion of \acp{DNS} through the Galaxy yields even deeper constraints on the \ac{SN} explosion of the progenitor stars, such the magnitude of \ac{SN} natal kicks and the amount of mass ejected by the exploding star~\citep{Wex2000,Willems2004,Wong2010,Andrews2015,Tauris2017,Andrews2019}. 
It is clear that some \ac{DNS} progenitors receive substantially weaker natal kicks at formation than is typical of isolated neutron stars~\citep{Willems2004a,Piran2005,Willems2006,Wong2010}, possibly due to stripping of the progenitor's envelope prior to \ac{SN}~\citep{Tauris2013,Tauris2015} or to an increased susceptibility to electron-capture \acp{SN} as opposed to standard iron core-collapse \acp{SN}~\citep{Podsiadlowski2004,vandenHeuvel2004,Schwab2010,Bray2016,Beniamini2016,Tauris2017}. 
On the other hand, certain Galactic \ac{DNS} systems, such as PSR\,B1534+12 and PSR\,B1913+16, are consistent with large natal kicks of ${\gtrsim 200~\mathrm{km\,s}^{-1}}$~\citep{Fryer1997,Wex2000,Wong2010,Tauris2017}. 
Many systems are less informative, and are observationally consistent with either large (${\gtrsim 100~\mathrm{km\,s^{-1}}}$) or small (${\lesssim 50~\mathrm{km\,s^{-1}}}$) natal kicks at formation~\citep[see][for a review]{Tauris2017}. 
Extragalactic information about \ac{DNS} natal kicks can also be gleaned from the multimessenger detection of \ac{DNS} mergers. 
However, GW170817's proximity to its host galaxy did not allow for strong constraints on the natal kick required to migrate the system to its merger location~\citep{GW170817_progenitor,Andreoni2017,Blanchard2017,Levan2017,Pan2017}. 

Both short \acp{GRB} we examine push toward high systemic velocities; across all halo masses, we find the posterior distributions for GRB~070809 and GRB~090515 to be \VsysNinetyGRBOne and \VsysNinetyGRBTwo at the $90\%$ credible level, respectively.
Figure~\ref{fig:weighted_offset} also shows the importance of accounting for the entire evolution of the host galaxy when interpreting systemic velocities from short \ac{GRB} offsets. 
In the context of highly offset \acp{GRB}, these systems can result from both \acp{DNS} that form early in the history of the galaxy, with long inspiral times and relatively low kicks that still allow them to explore the outer reaches of the lightweight galactic halo, as well as from \acp{DNS} that form at late times with shorter inspiral times that receive large enough kicks to escape the gravitational potential well of their host.

The systemic velocities alone are not sufficient to place constraints on \ac{SN} natal kicks. 
Commonly in the literature, constraints on systemic velocities are mistaken for constraints on \ac{SN} natal kicks, when in reality the post-\ac{SN} barycentric velocity of a compact binary is determined by the interplay of the pre-\ac{SN} galactic velocity, the \ac{SN} natal kick, the pre-\ac{SN} orbital separation, and the mass loss in the \ac{SN}. 
By marginalizing out other parameters that affect the post-\ac{SN} systemic velocity and inspiral time, we place the first constraints barring GW170817 solely on natal kicks using \ac{GRB} offsets. 

Large systemic velocities can result from tight pre-\ac{SN} orbital separations, though the pre-\ac{SN} separation is anticorrelated with the binary-inspiral time.\footnote{As a point of reference, a $1.4\,\mathrm{M}_\odot+1.4\,\mathrm{M}_\odot$ \ac{DNS} with an orbital separation of $1\,\mathrm{R}_{\odot}$ on a circular orbit will merge due to gravitational radiation in $\approx\,27~\mathrm{Myr}$ \citep{Peters1964}. } 
Since the hosts of both GRB~070809 and GRB~090515 have old stellar populations, the progenitors to the \acp{GRB} likely formed with pre-\ac{SN} orbital separations of ${\gtrsim 3\,\mathrm{R}_{\odot}}$ and relatively long inspiral times of ${\gtrsim 1~\mathrm{Gyr}}$, such that the contribution of the mass-loss kick to the systemic velocity is subdominant. 
\ac{SN} natal kicks are thus pushed to larger values to reconcile the systemic velocities necessary to produce the \ac{GRB} offsets. 
We find that highly offset short \acp{GRB}, particularly GRB~090515, necessitate large natal kicks of $\gtrsim 200~\mathrm{km\,s}^{-1}$ for most assumptions about halo  mass, and may indicate a formation scenario similar to Galactic \acp{DNS} with large inferred natal kicks, such as PSR\,B1913+16.

\subsection{Implications for Short \acp{GRB}}

Most short \acp{GRB} are detected at high redshifts for which we can only obtain a limited amount of information on the hosts; both short \acp{GRB} in this case study have redshifts of $z \gtrsim 0.4$. 
Therefore, we have devised a framework in which we make a number of empirically motivated assumptions in order to reverse-engineer galactic properties. 
We find that minor adjustments do not noticeably affect our results, so long as substantial amounts of star formation in the galactic hosts do not persist until the time of the \ac{GRB}. 
Our realization of the \ac{SMHM} relation, on the other hand, impacts our inference. 
As the halo mass increases and it becomes more difficult for \acp{DNS} to reach large offsets from their hosts (Figure~\ref{fig:weighted_offset}), systemic velocities and natal kicks must push to more extreme values to accommodate (see Figure~\ref{fig:kinematic} and Figure~\ref{fig:Vkick_constrains}). 
However, the majority of our key results are not significantly impacted by the dispersion of the \ac{SMHM} relation; 
GRB~090515 disfavors small natal kick velocities even for low halo masses (Figure~\ref{fig:Vkick_constrains}). 

It is possible for the association between a particular \ac{GRB} and its host to be incorrect. 
Based on the total stellar mass of nearby galaxies and the proximity of the \ac{GRB} to these galaxies, \citet{Fong2013a} calculated a $P_\mathrm{cc}$ of $3\%$ and $15\%$ for the hosts of GRB~070809 and GRB~090515, respectively.\footnote{The next most likely hosts for GRB~070809 and GRB~090515 have a $P_\mathrm{cc}$ of $\approx 10\%$ and $\approx 25\%$, respectively \citep{Berger2010}.} 
However, we find that these host associations lead to large, yet plausible, projected offset distributions, with \LargerOffsetsThanGRBOne of mergers at projected offsets greater than GRB~070809 and \LargerOffsetsThanGRBTwo greater than GRB~090515 in their respective hosts when assuming $\sigma_\mathrm{DM}=0$. 
As these \acp{GRB} are two of the more extreme examples of localized short \acp{GRB}, we find these host associations to be reasonable and explainable in the context of their \ac{DNS} progenitors. 
Determining whether these systems are indeed outliers and hint at an incorrect host association would require a population analysis with the full catalog of localized short \acp{GRB}. 

Even if the host galaxies are correctly identified, their evolutionary histories are more complicated than what is included in our simplified, semi-analytic galaxy models. 
We do not account for realistic cosmological environments, and in particular the possible impact of any major or minor galaxy mergers that may occur over the course of the Gyr binary-inspiral timescales.
Given that the host galaxies of \acp{GRB} are relatively distant, detailed information about their merger history is difficult to attain.
For galaxies with masses comparable to the hosts of GRB~070809 and GRB~090515, the merger rate is typically between $0.1$--$1\,\textrm{Gyr}^{-1}$ for mass-ratios $\gtrsim\,10^{-1}$~\citep{Rodriguez-Gomez2015}. 
Roughly $20\%$ of stellar mass is typically accreted from mergers, with the majority of mergers (especially major ones) occurring at high redshifts~\citep{Rodriguez-Gomez2016}.  
\ac{DM}-only simulations suggest that cosmological evolution does extend the radial distribution of compact binaries, but primarily those that already have relatively large systemic velocities~\citep{Kelley2010}. 
How much of this is caused by galaxy mergers as opposed to growth of galaxies over cosmic time, which we do account for, is unclear. 
Ultimately more modern, baryonic, and fully cosmological simulations should be performed to fully understand the effects on the distribution of compact binary mergers. 

Though follow-up of short \ac{GRB} counterparts provides limited information about \ac{DNS} systems relative to what can be attained for Galactic \acp{DNS}, such as precise proper motions and NS masses, short \acp{GRB} have the advantage of probing a broad spectrum of galaxy types over cosmic time. 
Furthermore, the observational biases obscuring these two \ac{DNS} populations are distinct, so any inferred constraints are complementary. 
Galactic \acp{DNS} are less likely to be detected if the binary has a very tight orbit, since these systems merge quicker and the large orbital acceleration leads to a fast-changing Doppler shift of the pulsar that smears the signal \citep[e.g.,][]{Tauris2017}. 
This could potentially bias the Galactic population to systems that have longer inspiral times and smaller post-\ac{SN} systemic velocities, since these are anticorrelated for tight binaries~\citep{Kalogera1996}. 
Short \acp{GRB} offer an independent population to explore that is not afflicted by the same selection effects inherent to the Galactic \ac{DNS} population. 

However, short \acp{GRB} have their own selection biases; for example, as \ac{DNS} mergers reach extreme offsets, the diffuse intergalactic medium will lead to dimmer optical afterglows~\citep{Paczynski1993,Meszaros1997,Sari1998}, potentially preventing the most highly offset \acp{GRB} from robust associations. 
We find that, for our two \ac{GRB} hosts, \FiveReffApprox of \ac{GRB} candidates have offsets exceeding $5\,R_\mathrm{e}$. 
This is lower than what is found from the observational sample in \citet{Fong2013a}, which finds $\approx 20\%$ of short \acp{GRB} merging at $\gtrsim 5\,R_\mathrm{e}$. 
The discrepancy between these offset distributions could be due to only considering short \acp{GRB} with massive elliptical hosts in this study, which will drive the offset distribution to lower values. 
Thus, from this analysis alone it is difficult to diagnose the selection effects possibly impinging the observational distribution of short \ac{GRB} offsets. 
A thorough examination, which properly accounts for a range of host galaxy properties and the fact that the majority of short \acp{GRB} occur in less massive, star-forming galaxies, is reserved for future work. 

Observations and subsequent localization of short \acp{GRB} can paint a complementary picture of \ac{DNS} formation channels. 
In this work, we focus on two exemplary highly offset short \acp{GRB}. 
To place comprehensive constraints on \ac{DNS} population properties, fully constrain selection effects, and examine deviations over cosmic time, this analysis can be applied to the full population of localized short \acp{GRB} with identified hosts as well as gravitational-wave mergers with optical counterparts. 

The galaxy models, tracer information, and population models used in this work are available on Zenodo~\citep{sgrb_progenitor_dataset}. 
The code used for the full analysis and plotting scripts for the figures in this paper are available on Github at \href{https://github.com/michaelzevin/kickIT}{github.com/michaelzevin/kickIT}.

\acknowledgments
The authors thank Claude-Andr\'e Faucher-Gigu\'ere, John Forbes, Joel Leja, Erica Nelson, and Enrico Ramirez-Ruiz for helpful discussions, as well as Chase Kimball for contributions to the code used in this analysis. 
The authors also thank the anonymous referee for helpful suggestions that improved this paper. 
Support for this work was partially provided by NASA through the NASA Hubble Fellowship grant HST-HF2-51474.001-A awarded by the Space Telescope Science Institute, which is operated by the Association of Universities for Research in Astronomy, Incorporated, under NASA contract NAS5-26555. 
M.Z. also appreciates financial support from the IDEAS Fellowship, a research traineeship program supported by the National Science Foundation under grant DGE-1450006. 
A.N. acknowledges support from the Henry Luce Foundation through a Graduate Fellowship in Physics and Astronomy.
W.F. acknowledges support by the National Science Foundation under grant Nos. AST-1814782 and AST-1909358. 
C.P.L.B. is supported by the CIERA Board of Visitors Research Professorship.
V.K. is supported by a CIFAR G+EU Fellowship and Northwestern University. 
The majority of our analysis was performed using the computational resources of the Quest high performance computing facility at Northwestern University, which is jointly supported by the Office of the Provost, the Office for Research, and Northwestern University Information Technology.

\software{\texttt{Astropy}~\citep{TheAstropyCollaboration2013,TheAstropyCollaboration2018}, \texttt{COSMIC}~\citep{Breivik2020}, \texttt{galpy}~\citep{galpy}, \texttt{IPython}~\citep{ipython}, \texttt{matplotlib}~\citep{matplotlib}, \texttt{Numpy}~\citep{numpy,numpy2,numpy3}, \texttt{pandas}~\citep{pandas}, \texttt{Prospector}~\citep{Leja2016}, \texttt{Scipy}~\citep{scipy}.}

\pagebreak
\bibliographystyle{aasjournal}
\bibliography{library}

\appendix

\section{Time-dependent Galactic Models}\label{app:hostgal}

With \ac{SFH}s in hand for each galaxy model, we reverse-engineer the short \ac{GRB} host galaxies by evolving the observed stellar mass backwards in time as
\begin{equation}
    M_{\star}(t) = M_{\star}^{0} - \int_{t}^{t_\mathrm{GRB}} \dot{M}_{\star}(t') \,\mathrm{d}t',
\end{equation}
where $\dot{M}_{\star}(t)$ is the \ac{SFR} at time $t$. 
We integrate this backwards until $M_{\star} = 0$ to determine the approximate formation time of the galaxy. 

\begin{figure*}[t!]\label{fig:galaxy_model}
\includegraphics[width=1.0\textwidth]{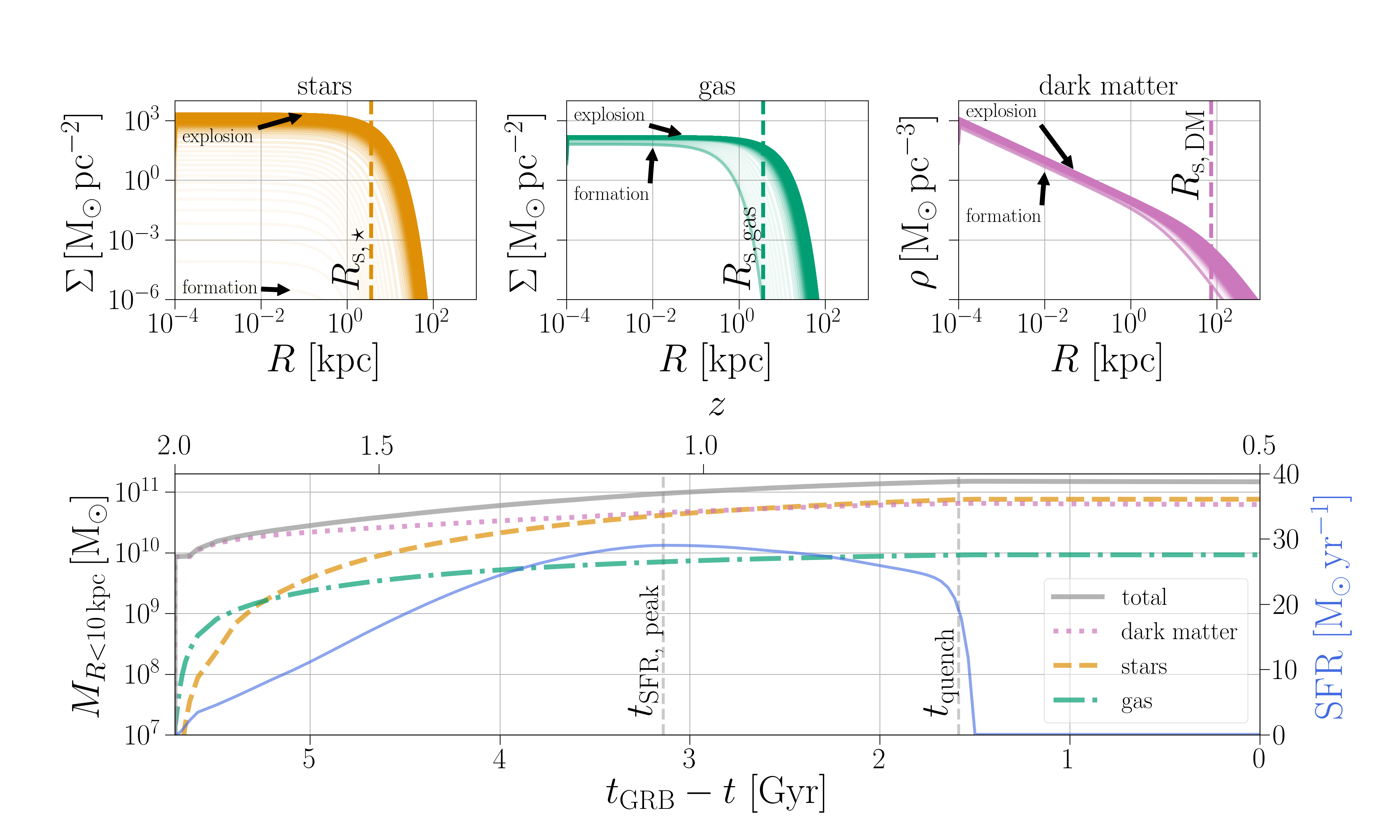}
\caption{Galaxy model for the probable host of GRB~070809. 
The top row shows surface densities of the stellar and gas components, which follow an exponential disk profile, and the volume density of the \ac{DM} component, which follows an \ac{NFW} profile. 
Lines of increasing opacity show the density profile at different redshifts, with the most opaque line for $z=z^{0}$; arrows point to the initial density and final density of the galactic components. 
Scale radii for each component at the time of the \ac{GRB} are also shown with dashed lines. 
The bottom panel shows the evolution of the component masses (left axis) and the \ac{SFR} (right axis) as a function of time. 
For each component, we show the mass enclosed within a sphere of radius 10\,kpc, as the majority of the mass in stars and gas (and therefore the birth location of most tracer particles) falls within this distance. 
The time of peak star formation and the quenching time, two of the key parameters used in our modeling, are marked with dashed gray lines. 
}
\end{figure*}

With a stellar mass and \ac{SFR} determined at each point in time, we construct a three-component, time-dependent model for the galactic potential that accounts for gas, stars, and \ac{DM}. 
At each step in time, the gas mass $M_\mathrm{gas}(t)$ is determined using the fits from \cite{Peeples2014}, assuming a $50\%$ warm gas fraction~\citep{Oey2007}.
The \ac{DM} halo mass $M_\mathrm{DM}(t)$ is determined using the \ac{SMHM} relation from \cite{Moster2013}. 
We model multiple realizations for each galaxy with different deviations from the median of the \ac{SMHM} relation and marginalize over this variance in our uncertainty estimates. 
The \ac{SFR} radial distribution is assumed to follow an exponential disk, with the characteristic scale radius of the star-forming disk at each point in time $\mathcal{R}_\mathrm{s,\star}(t)$ determined as in \cite{Nelson2016}. 
We account for dispersion in this relation by adjusting the scale radius at each time by the fractional difference between the observed effective radius and the scale radius calculated at the time of the GRB: 
\begin{equation}
    \mathcal{R}_{\mathrm{s},\star}(t) = 
    \frac{R_\mathrm{e}^{0}}{\bar{\mathcal{R}}_{\mathrm{s},\star}(t=t_\mathrm{GRB})} \bar{\mathcal{R}}_{\mathrm{s},\star}(t)
    ,  
\end{equation}
where $\bar{\mathcal{R}}_{\mathrm{s},\star}(t)$ is calculated using the fits from \citet{Nelson2016}. 
The \ac{DM} scale radius $\mathcal{R}_{\mathrm{s}, \mathrm{DM}}(t)$ is calculated assuming the \ac{DM} is distributed in a \ac{NFW} profile~\citep{Navarro1997}, with a concentration parameter determined from \citet{Klypin2016}. 
Figure~\ref{fig:galaxy_model} shows the variation of radial density profiles for each galactic component as the galaxy grows, as well as the evolution of the galaxy mass over cosmic time. 

Our three-component model for the galactic potential at each point in time is
\begin{equation}
    \Phi(t) = \Phi_{\star}(t) + \Phi_\mathrm{gas}(t) + \Phi_\mathrm{DM}(t),
\end{equation}
where $\Phi_{\star}$ and $\Phi_\mathrm{gas}$ are assumed to follow a double-exponential density profile with scale height $\mathcal{R}_{\mathrm{\star},z}(t) = 0.1 \, \mathcal{R}_\mathrm{s,\star}(t)$ \citep[e.g.,][]{Kregel2002}, and $\Phi_\mathrm{DM}$ follows an \ac{NFW} profile, which is parameterized by $M_\mathrm{DM}(t)$ and $\mathcal{R}_\mathrm{s,\,DM}(t)$.\footnote{For star forming galaxies, observations typically give $\mathcal{R}_{\mathrm{\star},z} / \mathcal{R}_\mathrm{s,\star} \approx 0.15$--$0.20$ \citep[e.g.,][]{Bottema1993,deGrijs1996,Bizyaev2009} instead of our adopted value of $0.1$.  However, the youngest stars and thus ongoing star formation tend to be more narrowly distributed around the midplane \citep[e.g.,][]{Seth2005}.  For simplicity and because, at each epoch, the disk potential is likely most important for the currently forming stars, we adopt the given value.} 
For $\Phi_\mathrm{gas}$ and $\Phi_\mathrm{DM}$, the potential only accounts for the distribution of material at a given point in time. 
However, as stellar mass is built up from the preexisting distribution of star-forming gas, $\Phi_{\star}(t)$ is determined by the differential star formation profiles for all times preceding $t$:
\begin{equation}
    \triangledown^2 \Phi_\star(t) = 4 \pi G
    \int_{t_\mathrm{form}}^{t}\mathcal{A}(t')\exp\left({-\frac{r}{\mathcal{R}_{\mathrm{s},\star}(t')} - \frac{|z|}{\mathcal{R}_{z,\star}(t')}}\right) \,\mathrm{d}t',
\end{equation}
where the integrand is the cumulative double-exponential density profile at time $t$ and the normalization amplitude is proportional to the differential stellar mass at each point in time,
\begin{equation}
    \mathcal{A}(t) = \frac{\dot{M}_{\star}(t)} {4 \pi \mathcal{R}_{z,\star}(t) \mathcal{R}_{\mathrm{s},\star}(t)^2}.
\end{equation}

In practice, we discretely sample points in time between $t_\mathrm{form}$ and $t_\mathrm{GRB}$ to approximate the continuous evolution of the galaxy over cosmic time. 
To ensure a fine resolution in time, we require that each time step not exceed $0.1~\mathrm{Gyr}$ and that the fractional change in stellar mass at each step not exceed $1\%$. 
For the \ac{GRB} host galaxies we examine, this leads to $\approx\,150$ potential models between the formation time of the galaxy and the time of the \ac{GRB}.

\section{Seeding Tracer Particles}\label{app:tracers}

Given our galactic models, the probability of a system being born at time $t_\mathrm{birth}$ is 
\begin{equation}
    p(t=t_\mathrm{birth}) = \frac{M_\mathrm{gas}(t) \dot{\mu}_\star(t)}{\int_{t_0}^{t_\mathrm{GRB}} M_\mathrm{gas}(t') \dot{\mu}_\star(t') \,\mathrm{d}t},
\end{equation}
where $\dot{\mu}_\star(t)$ is the specific \ac{SFR} at time $t$. 
The inspiral time of the kinematic tracers is assumed to be \mbox{$t_\mathrm{insp}^\mathrm{kin} = t_\mathrm{GRB} - t_\mathrm{birth}$}, such that the \ac{DNS} mergers occur at the correct redshift of the \ac{GRB}. 

Each tracer particle $i$ is distributed in the plane of the galaxy according to the gas-density profile at time $t_\mathrm{birth}$, and assigned an initial circular velocity in the galaxy according to the galactic potential at that time. 
We then apply a randomly oriented post-\ac{SN} systemic velocity, $V_\mathrm{sys}^\mathrm{kin}$, to the tracer, sampled uniformly in magnitude from $[0, 1000]~\mathrm{km\,s}^{-1}$, which explores the possible range of post-\ac{SN} systemic velocities due to the \ac{SN} that formed the second neuron star.\footnote{The systemic velocities are probability weighted in post-processing (see Section \ref{subsec:bayesian}).}

\section{Posterior Distributions for Population Parameters}\label{app:bayesian}

Using Bayes' theorem, we can rewrite the joint posterior distribution on $\vec{\Theta}_\mathrm{kin}$ as
\begin{equation}
    p(\vec{\Theta}_\mathrm{kin} | \vec{\Theta}_\mathrm{obs}) 
    = \frac{p(\vec{\Theta}_\mathrm{obs} | \vec{\Theta}_\mathrm{kin}) p(\vec{\Theta}_\mathrm{kin})}{ p(\vec{\Theta}_\mathrm{obs})}
\end{equation}
where $p(\vec{\Theta}_\mathrm{obs} | \vec{\Theta}_\mathrm{kin})$ is the likelihood of recovering the observed offset given our kinematic modeling, $p(\vec{\Theta}_\mathrm{kin})$ is the prior on our kinematic parameters, and $p(\vec{\Theta}_\mathrm{obs})$ is a normalization constant. 
The projected offset and redshift of merger are the variables of interest from our kinematic modeling, where the projected physical offset is dependent on the redshift of merger. 
We expand the likelihood in terms of these variables: 
\begin{equation}
    p(\vec{\Theta}_\mathrm{kin} | \vec{\Theta}_\mathrm{obs}) 
    \propto \int p(\vec{\Theta}_\mathrm{obs} | R_\mathrm{off}, z)  p(R_\mathrm{off}, z | V_\mathrm{sys}, R_\mathrm{birth}, z_\mathrm{birth}) p(V_\mathrm{sys}, R_\mathrm{birth}, z_\mathrm{birth})
    \,\mathrm{d}R_\mathrm{off}\, \mathrm{d}z. 
\end{equation}
The first expression in the integrand, $p(\vec{\Theta}_\mathrm{obs} | R_\mathrm{off}, z)$, is the observational weighting of the likelihood. 
We enforce that $z = z^{0}$, and assume the uncertainty in the offset measurement is Gaussian distributed, such that
\begin{equation}
\begin{aligned}
    w_\mathrm{obs} &\equiv p(\vec{\Theta}_\mathrm{obs} | R_\mathrm{off}, z) \\
    &= \frac{1}{\sqrt{2 \pi \sigma_{R_\mathrm{off}^{0}}^2}} 
    \exp \left( -\frac{(R_\mathrm{off} - R_\mathrm{off}^{0})^2}{2 \sigma_{R_\mathrm{off}^{0}}^2} \right) \delta(z - z^{0}). 
\end{aligned}
\end{equation}
The second expression in the likelihood, $w_\mathrm{kin} \equiv p(R_\mathrm{off},z$ $| V_\mathrm{sys}, R_\mathrm{birth}, z_\mathrm{birth})$, is the result of our kinematic modeling, where $R_\mathrm{birth}$ and $z_\mathrm{birth}$ are determined by the galaxy model described in Section \ref{subsec:galmodel}. 
Since we fix the merger redshift to be the observed $z^{0}$, the birth redshift solely determines the inspiral time of the binary, and we can replace $z_\mathrm{birth}$ with $t_\mathrm{insp} = \mathcal{T}(z^0) -  \mathcal{T}(z_\mathrm{birth})$ where $\mathcal{T}$ is the transformation from redshift to proper time. 

Rewriting the prior term as $p(V_\mathrm{sys}, R_\mathrm{birth}, t_\mathrm{insp})$, we separate the components of the prior that come from our galaxy model from those that come from our population model as
\begin{equation}
\begin{aligned}
    p(V_\mathrm{sys}, &R_\mathrm{birth}, t_\mathrm{insp}) 
    = p_\mathrm{pop}(V_\mathrm{sys}, t_\mathrm{insp})  p_\mathrm{gal}(V_\mathrm{sys}) p_\mathrm{gal}(R_\mathrm{birth} | t_\mathrm{insp}) p_\mathrm{gal}(t_\mathrm{insp}),
\end{aligned}
\end{equation}
where $w_\mathrm{pop} \equiv p_\mathrm{pop}(V_\mathrm{sys}, t_\mathrm{insp})$ is the prior on the joint systemic velocity and inspiral time distribution from \ac{DNS} population modeling as described in Section \ref{subsec:popmodel}, $p_\mathrm{gal}(V_\mathrm{sys})$ is sampled uniformly (and thus does not affect the posterior), $p_\mathrm{gal}(R_\mathrm{birth} | t_\mathrm{insp})$ is the exponential radial profile at a given redshift from which we population tracers, and $p_\mathrm{gal}(t_\mathrm{insp})$ is the injected inspiral time of tracer particles, which is determined from the \ac{SFH} of the galaxy model. 
We define $w_\mathrm{gal} \equiv p_\mathrm{gal}(V_\mathrm{sys}) p_\mathrm{gal}(R_\mathrm{birth} | t_\mathrm{insp}) p_\mathrm{gal}(t_\mathrm{insp})$  for simplicity. 
The posterior distribution on $\vec{\Theta}_\mathrm{kin}$ given the observations is therefore 
\begin{equation}
    p(\vec{\Theta}_\mathrm{kin} | \vec{\Theta}_\mathrm{obs}) \propto \int w_\mathrm{obs} \times w_\mathrm{kin} \times w_\mathrm{gal} \times  w_\mathrm{pop}\,\mathrm{d}R_\mathrm{off}\,\mathrm{d}z, 
\end{equation}
which is Eq.~\eqref{eq:kin_posterior}.

To back out constraints on the \ac{DNS} progenitor parameters, we again invoke Bayes' theorem to rewrite $w_\mathrm{pop}$ as $p(\vec{\Theta}_\mathrm{kin}|\vec{\Theta}_\mathrm{pop}) p(\vec{\Theta}_\mathrm{pop}) / p(\vec{\Theta}_\mathrm{pop} | \vec{\Theta}_\mathrm{kin})$. 
Multiplying through by $p(\vec{\Theta}_\mathrm{pop} | \vec{\Theta}_\mathrm{kin})$ and marginalizing over the kinematic parameters $\vec{\Theta}_\mathrm{kin}$, we get
\begin{equation}
    \int p(\vec{\Theta}_\mathrm{pop} | \vec{\Theta}_\mathrm{kin}) p(\vec{\Theta}_\mathrm{kin} | \vec{\Theta}_\mathrm{obs}) \,\mathrm{d}\vec{\Theta}_\mathrm{kin} \propto \int w_\mathrm{obs} w_\mathrm{kin} w_\mathrm{gal} p(\vec{\Theta}_\mathrm{kin}|\vec{\Theta}_\mathrm{pop}) p(\vec{\Theta}_\mathrm{pop}) \,\mathrm{d}R_\mathrm{off} \,\mathrm{d}z \,\mathrm{d}\vec{\Theta}_\mathrm{kin}. 
\end{equation}
By noting that the birth location in the galaxy, $R_\mathrm{birth}$, is independent of the population properties, the posterior distribution for the population parameters can be condensed as 
\begin{equation}
    p(\vec{\Theta}_\mathrm{pop} | \vec{\Theta}_\mathrm{obs}) 
    \propto \int w_\mathrm{obs} w_\mathrm{kin} w_\mathrm{gal} p(V_\mathrm{sys}, t_\mathrm{insp} | \vec{\Theta}_\mathrm{pop}) p(\vec{\Theta}_\mathrm{pop}) \,\mathrm{d}R_\mathrm{off} \,\mathrm{d}z \,\mathrm{d}V_\mathrm{sys} \,\mathrm{d}t_\mathrm{insp}. 
\end{equation}

\end{document}